\documentclass[journal,twoside]{paper}
\usepackage{cite}
\usepackage{multirow}
\usepackage[table,xcdraw]{xcolor}
\usepackage{amsmath,amssymb,amsfonts}
\usepackage{algorithmic}
\usepackage{stfloats}
\usepackage{graphicx}
\usepackage{subfigure}
\usepackage{textcomp}
\def\BibTeX{{\rm B\kern-.05em{\sc i\kern-.025em b}\kern-.08em
    T\kern-.1667em\lower.7ex\hbox{E}\kern-.125emX}}
\usepackage{tikz}
\usetikzlibrary{shapes.geometric, arrows}

\begin{document}
\title{Joint Optical Neuroimaging Denoising with Semantic Tasks}
%\author{First A. Author, \IEEEmembership{Fellow, IEEE}, Second B. Author, and Third C. Author, Jr., \IEEEmembership{Member, IEEE}
\author{Tianfang Zhu, Yue Guan$\star$%%$^\dagger$, Tianfang Zhu$^\dagger$, Zhangheng Ding$^\dagger$, Jing Yuan, Hui Gong, and Anan Li$^\star$
%  \thanks{$\dagger$: authors with equal contribution.}
%\thanks{$\star$: corresponding authors.}
\thanks{This work was supported by the National Natural Science Foundation of China (Grant Nos. 81827901, 61890954, 61890950).}
%\thanks{The next few paragraphs should contain 
%the authors' current affiliations, including current address and e-mail. For 
%example, F. A. Author is with the National Institute of Standards and 
%Technology, Boulder, CO 80305 USA (e-mail: author@boulder.nist.gov). }
\thanks{The author is with Huazhong University of Science and Technology, Wuhan, China (e-mail: yguan@hust.edu.cn).}
}

\maketitle

\begin{abstract}
%The difficulty of micro-image denoising is how to get stronger denoising capability in the absence of ground truth.
%In this paper, a self-supervised denoising method for end-to-end training combined with follow-up tasks is proposed.
%By linking Noise2Void with downstream models, denoised model can be benefited from subsequent application and further improve effectiveness of itself.
%When matching models are difficult to be obtained, we provide an overall unsupervised learning scheme to alternative: \
%Determine an expected clean style according to a similar segmentation model,
%and then drive the denoiser closer to it to a certain extent.
%Experiments shows that the method presented in this paper has a further denoising effect
%on both light and electron microscope data.
%Because downstream tasks are taken into account,
%denoised images get higher scores in subsequent segmentation and recognition tasks.
%
  Optical neuroimaging is a vital tool for understanding the brain structure and the connection between regions and nuclei.
  However, the image noise introduced in the sample preparation and the imaging system hinders the 
  extraction of the possible knowlege from the dataset,
  thus denoising for the optical neuroimaging is usually necessary.
  The supervised denoisng methods often outperform the unsupervised ones,
  but the training of the supervised denoising models needs the corresponding clean labels,
  which is not always avaiable due to the high labeling cost.
  On the other hand, those semantic labels, such as the located soma positions, the reconstructed neuronal fibers, and the nuclei segmentation result,
  are generally available and accumulated from everyday  neuroscience research.
  This work connects a supervised denoising and a semantic segmentation model together to form a end-to-end model,
  which can make use of the semantic labels while still provides a denoised image as an intermediate product.
  We use both the supervised and the self-supervised models for the denoising
  and introduce a new cost term for the joint denoising and the segmentation setup.
  We test the proposed approach on both the synthetic data and the real-world data,
  including the optical neuroimaing dataset and the electron microscope dataset.
  The result shows that the joint denoising result outperforms the one using the denoising method alone
  and the joint model benefits the segmentation and other downstream task as well.
\end{abstract}

%\begin{IEEEkeywords}
%  Denoising, Neuroimaging
%%Enter key words or phrases in alphabetical 
%%order, separated by commas. For a list of suggested keywords, send a blank 
%%e-mail to keywords@ieee.org or visit \underline
%%{http://www.ieee.org/organizations/pubs/ani\_prod/keywrd98.txt}
%\end{IEEEkeywords}

\section{Introduction}
\label{sec:introduction}
Optical neuroimaging at the mesoscopic scale is an important approach
to understand the brain structures
and the connection between these structures~\cite{hama2015scales}. %cite some review
It balances the resolution and the scope of the imaging process
and it may be one of few candidates for a whole brain imaging
at the single neuron scale~\cite{hildebrand2017whole,gao2019cortical,power2017sources}. % cite some review
The optical neuroimaging process usually starts after the time consuming sample preparation
and fixing process~\cite{veloo2014the}.
There are lots of sample preparation and staining methods depolyed for the optical neuroimaging,
including the Nissl staining~\cite{lin2019a,carriel2017staining}
and the labeling by a large family of green fluorescent proteins~(GFP)~\cite{rodriguez2017the,emanuel2017high}. % cite for nissle and gfp
Image acqusition methods,
such as the serial-two-photon~(STP)~\cite{ragan2012serial,amato2016whole},
the structured illumination~(SI)~\cite{hell2015the},
the time delay integration~(TDI)~\cite{lepage2009time}
and others form the basis
for the mesoscopic scale optical neuroimaging system. % each method gets a citation.
The image acqusition systems,
such as the micro optical sectioning tomograph~(MOST)~\cite{li2010micro}
and MouseLight~\cite{economo2016a,tasic2018shared,winnubst2019reconstruction},
integrate these processes to produce images showing the whole brain morphological structure.
Besides the ever improving imaging process,
it is inevitable to introduce the image noise
during the sample preparation
and the imaging process~\cite{jeong2020sparse,pizurica2006a,gong2020deep}.
Thus the image denoising is necessary to improve the quality of the dataset
and more importantly, it benefits the downstream processing and analysis tasks,
such as the soma positioning~\cite{parekh2015the,rees2017weighing},
the brain region segmentation~\cite{fischl2002whole,xu2020automated},
and the registration to the reference atlas~\cite{avants2011a,dalca2019unsupervised,duan2019adversarial}.

% say sth about denoise in general
There are extensive researches on the image denoising topic.
Several 2D image filters are developed for the denoising with the assumption
that the true pixel can be recovered
by constructing a mapping between the pixel and its neighbour pixels~\cite{tomasi1998bilateral}.
This denoising method based on the local information is extended both
in the spatial domain and the transformation domain.
The classic denoising methods based on the local information
include more filters~\cite{buades2005a,yu2002speckle}, %cite a filter
methods based on the nonlinear total variation~\cite{rudin1992nonlinear},
and others~\cite{milanfar2013a}.
The denoising methods that use the local information
while apply the denoising in the transformation domain
include the collaborative filtering in the transform-domain~\cite{dabov2007image},
, the sparse reconstruction based methods~\cite{mairal2009non}
, and other methods~\cite{katkovnik2017complex,li2019improved}. % cite a filter
The classic approach also seeks to use the global information
and the semantic information for the image denoising,
this includes the work in~\cite{hosseini2018semantic,lu2017learning,ding2019semantic}% add citation.
As the fast developing deep learning based methods bring the cutting-edge performance
on the visual tasks, it is also used for the image denoising.
Some works try to use the classic filter based method
with the deep learning model~\cite{burger2012image},
while others make use of the larger representation capacity of the deeper network
to achieve a better performance~\cite{xie2012image,mao2016image}.
These studies evaluate the performance of the deep learning based denoising methods
not only for the Gaussian noise but also for other noise distributions~\cite{zhang2017beyond,chang2020two}.
In recent years, self-supervised denoising,
where a supervised model is trained without the corresponding clean images,
brings a new possibility for the image denoising
when the clean label images are not available~\cite{krull2019noise2void}.
An experience learned from these studies is that
the supervised denoising methods are usually superior to the self-supervised methods
due to the fact that the denoising models are trained to map the input to the output
with a similar distribution of the clean images.
This leads to belief that when it's possible,
we may be better to bring more labeled data into the denoising model
to produce better results.
These additional labels may help to improve the performance of the supervised denoising models
and the self-supervised denoising models,
while the latter one is the common case in the optical neuroimaging,
where there are usually no clean labeled images available.

The image denoising is usually the first step among the optical mesoscopic image analysis steps and
there is usually no clean images available for the denoising model training.
On the other hand, when the image processing and analysis pipeline
has extracted the information from the image dataset,
the manual annotations usually become available as the clean labels for these tasks.
Such tasks may include the soma detection,
the brain region segmentation, and the neuronal fiber annotation.
It is common for the optical neuroimaging system to capture the image
and let the downstream image processing and analysis pipeline
to generate the neuroscience knowledge from the image dataset.
Thus it may be beneficial to join the tasks of the early
and the later stage of the optical neuroimaging data processsing pipeline
in order to use the annotation from the later stage.
This joint learning approach may improve both
the performance of the models used in the early stage and the later stage.
There are several works in variety of applications
trying to guide the early stage tasks using the downstreaming cost
to improve the performance of the early stage tasks, such as denoising.
Early works, such as~\cite{johnson2016perceptual} for the style transfer,
already use a two stages structure,
where the first stage model is the transform model to generated the output in the new style
and the second stage model is used to introduce the perceptual cost
to drive the output from the first stage model to the target style.
In the application such as the object recognition
in the very low resolution images,~\cite{wang2016studying}
tries to first use a super-resolution network
to convert the low resoltuion images into the higher resolution images
and then use the second network for the recognition task.
The traing of the super-resolution network is governed
by the recognition network to improve its performance.
A similar work investigate the image denoising
with the help of the high-level task is discussed in~\cite{liu2018when}.
The work uses the 3-scale UNet model~\cite{ronneberger2015} as the denoising model
and an image classification and a semantic segmentation model as the high-level task.
The high-level task model is first trained on the noisy dataset
and then it is connected to the first stage to drive the denoising trainng
while the weights of the high-level task models are fixed.
This approach is similar to our work described here,
where we go beyond the UNet model for the denoising model backbone
and discuss both the model-supervised and the label-supervised high-level task.
We also provide an empirical study of the application  for the optical neuroimaging dataset and the electrical microscope~(EM) image dataset.

% TODO: say sth about the unet and noise2void
Even with the help of the downstream analysis task,
it's still vital to understand
the representation capacity and the regularization of the denoising model.
The supervised denoising models, such as UNet~\cite{ronneberger2015},
was originally introduced as a segmentation model
and the network is usually modelling the noise residual when used as the denoising model.
However, UNet is a supervised model and provides little benefit when the labels are not available.
By introducing the self-supervised training, noise2noise model~\cite{lehtinen2018noise2noise}
allows the training without the labels but with only the paired noisy images
and in fact uses the paired noisy image as a regularization.
The noise2void model~\cite{krull2019noise2void}
introduces an even stronger regularization by assuming
taht the densities of local pixels  are similar with each other.
Thus we prefer the strong regularization in our proposed model given the high representation capacity
of general deep network models.

In summary, the contribution of this paper is two-fold:
\begin{enumerate}
\item We improve the optical mesoscopic neuroimaging denoising by joint learning the denoising and the semantic task,
  by cascading the unsupervised Noise2void model and U-Net segmentation model.
  It not only enhances the visual denoising effect, but also achieves higher performance in downstream tasks.
\item For the downstream tasks where the labels are not available for supervised training,
  we propose an  unsupervised alternative to the optical neuroimaging dataset
  to ensure that the denoiser can also benefit from the downstream knowledge.
\end{enumerate}

The paper presents the proposed model as follows.
Sec.~\ref{subsec:noise2void} introduces the modified noise2void denoising model;
Sec.~\ref{subsec:se2e} presents the supervised training and
Sec.~\ref{subsec:ue2e} presents the unsupervised training approach.
In the experiment, we present the result for the EM soma segmentation result
in Sec.~\ref{subsec:somaseg}
and the result for the optical mesoscopic plaques segmentation in Sec.~\ref{subsec:plagseg}.
The conclusion and the future work is discussed in Sec.~\ref{sec:conclusion}.

% done: fronzen  parameters -> pretrained model
% TODO: mitochondrion -> plaque
\begin{figure}[h]
\centerline{\includegraphics[width=9cm]{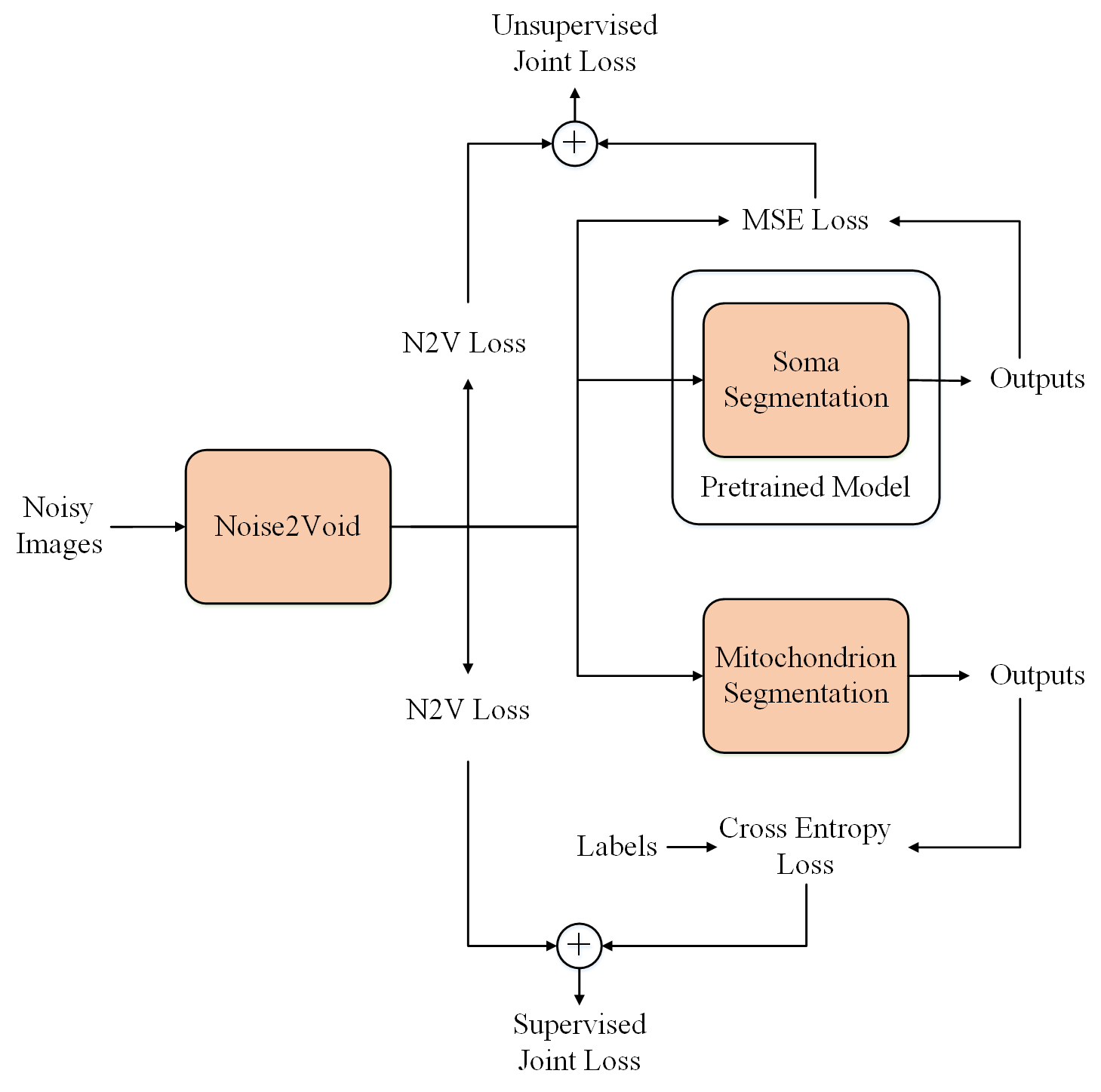}}
\caption{Joint optical mesoscopic neuroimaging denoising with semantic tasks.
  The denoising model is the modified noise2void model.
  The semantic task is either soma segmentation in one application
  and plaque segmentation in the other.
  When the labels are not available, a pretrained model is used.
  Otherwise, the semantic labels are used to drive both the denoising model and the segmentation model.
}
\label{fig:e2e pipeline}
\end{figure}

\section{Method}
\label{sec:method}
% TODO: intro and notation
In this paper, an end-to-end deep learning network from the image denoising to the downstreaming semantic tasks is constructed
to further improve the performance of the denoising model.
As shown in Fig.~\ref{fig:e2e pipeline}, the end-to-end model consists of two parts: the denoiser and the semantic task model.
In our case, the semantic tasks here are the mitochondrion segmentation for the EM image dataset and the $A\beta$ plaque recognition for the optical imaging dataset.
According to whether the follow-up model requires supervised training, 
Thus we divide the whole method into supervised end-to-end training and unsupervised end-to-end training, which are different in implementation and effect.
In order to reduce the limitation of clean labels on the denoising of microscopic image, 
the same self-supervised model Noise2Void~\cite{n2v} is used in the two training strategies.
The following is a detailed description of these three parts.

\subsection{Noise2Void Denoiser}
\label{subsec:noise2void}
% N2V is a self-supervised training method
% TODO: add an illustration to show what is the mask?
Recent studies show that the denoising model can be trained without clean images. 
For instance, Noise2Noise~\cite{n2n} method can train a pair of noise data to get the same excellent denoising model. 
However, it is still difficult to obtain independent imaging results of the same scene for obtained microscopic images.

Noise2Void~\cite{n2v} method goes further, trying to train the denoising model with a single noise image. 
It assumes that the signal in the image is continuous while the noise is independent, so the center pixel can be estimated from domain pixels.

Specifically, N2V proposes a training method called blind-spot network. 
Firstly, a series of small patches (64×64 by default) are extracted from the original images, 
then the center pixel is taken as the target, 
while the neighborhood pixels are used as the input during training.
In order to avoid learning identity mapping, 
N2V uses masking scheme, which randomly replaces the center pixel with another surrounding one.
The optimization formula is as follows in Equ.~\eqref{equ:opt}
\begin{equation}
  \underset{\boldsymbol{\theta}}{\arg \min } \sum_{j} \sum_{i} L\left(f\left(\tilde{\boldsymbol{x}}_{\mathrm{RF}(i)}^{j} ; \boldsymbol{\theta}\right), \boldsymbol{x}_{i}^{j}\right)
  \label{equ:opt}
\end{equation}
where $\boldsymbol{x}_{i}^{j}$ is the target pixel, $\tilde{\boldsymbol{x}}_{\mathrm{RF}(i)}^{j}$ is a patch around pixel $i$ and $\theta$ represents the parameters of the network model.
Here, the standard MSE loss is usually used,
which is only calculated at the replacement pixel as shown in Equ.~\eqref{equ:mseloss}
\begin{equation}
  L\left(\hat{s}_{i}^{j}, s_{i}^{j}\right)=\left(\hat{s}_{i}^{j}-s_{i}^{j}\right)^{2}
  \label{equ:mseloss}
\end{equation}

\begin{figure}[htbp]
\centerline{\includegraphics[width=0.3\textwidth]{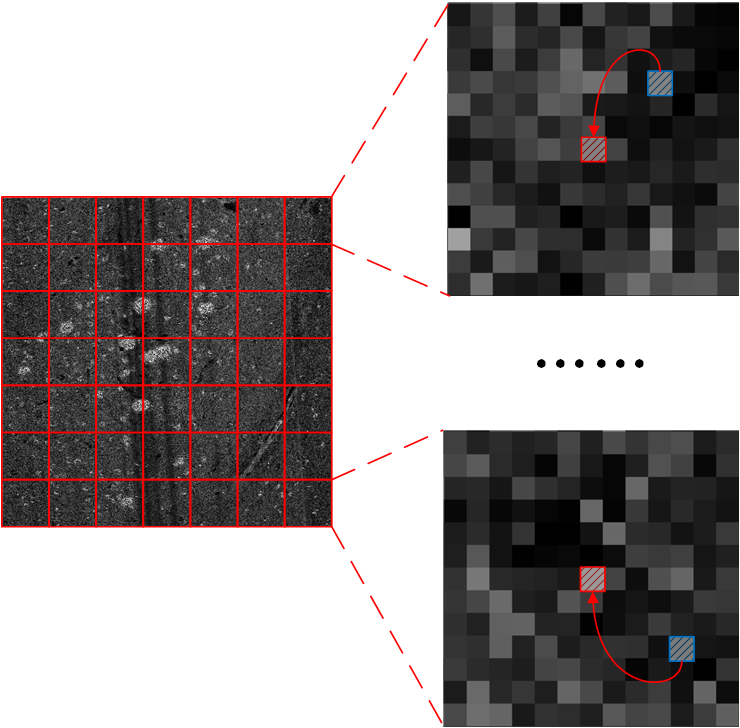}}
\caption{The pixel masking scheme of our end-to-end network}
\label{fig:mask}
\end{figure}

For the obvious advantages of microscopic images, 
we use N2V method as the common denoiser of end-to-end model. 
Considering that these patches may not contain the subjects of downstream tasks, including recognition and segmentation, 
we input the original size images directly
(see Fig.~\ref{fig:mask}). 
We divide the original image into several sub regions 
and then replace the center pixel randomly for each sub region, the loss is calculated of all the replaced pixels in the image.

\subsection{Supervised end-to-end training}
\label{subsec:se2e}
If the training data of the high-level semantic task is available,
a supervised model can be directly connected after the denoising model
to form an supervised end-to-end training network.

% model
Here we take a join segmentation task as an example and the training strategy is described in detail below:
Denoising model is an modified U-Net structure.
The activation function is replaced by leaky rectified linear unit~(Leaky ReLU),
and the batch normalization layers are added before each activation function.
Noise2Void adopts the MSE loss.
In order to continuously segment after denoising,
we directly calculates MSE loss by randomly replacing more pixels in full-size images.
The distribution of biological structures in mesoscopic images is relatively sparse,
training through larger size images is also helpful to capture global features.
Denoising loss is recorded as $l_1$.
Segmentation model has the same structure of denoiser,
and binary cross entropy loss is used for pixel level classification.
In the whole end-to-end training, the segmentation model only provides the knowledge of downstreaming tasks
to further improve the performance of the denoising model,
so it is necessary to avoid the situation that segmentation loss is so low
that make denoiser could not improve due to the over-powerful segmentation model.
% If looking forward to computer understands the segmentation task completely by itself,
If training the segmentation task without regularization,
it will add a lot of uncertainty to the whole end-to-end training process.
Therefore, a compromise scheme is adopted in this paper.
Firstly, a weaker segmentation model is trained separately,
which can be done by setting fewer training steps,
afterwards the model can coordinately participate in the end-to-end network
and update its own parameters.
The partition loss is denoted as $l_2$.
The combined loss can be obtained through a weighted summation of the denoising loss and the segmentation loss as shown in Equ.~\eqref{equ:weightedloss}.
\begin{equation}
l = w_1 l_1 + w_2  l_2  
  \label{equ:weightedloss}
  \end{equation}
We can set different weights to control the loss distribution between the denoising loss and the segmentation loss.
A high weight value for the segmentation loss will sometimes cause visual incongruence of denoised images.

\subsection{Unsupervised end-to-end training}
\label{subsec:ue2e}
% intro
The supervised end-to-end network can be directly constructed
when the training dataset for downstreaming task is available.
However, for real world optical image processing task,
it is difficult to obtain the corresponding high-level semantic labels for a noisy image dataset.
On one hand, the low signal-noise ratio  will greatly increase the difficulty of manual marking;
on the other hand, these images may be regarded as low quality data and then discarded directly in vein.
Here we present an unsupervised end-to-end alternative scheme that does not require clean images or the target semantic labels.
We observe that there is a strong similarity among some high-level semantic tasks of microscopic images in neuroscience research.

% TODO: add model output vs ground truth.
\begin{figure}[h]
\centerline{\includegraphics[width=0.45\textwidth]{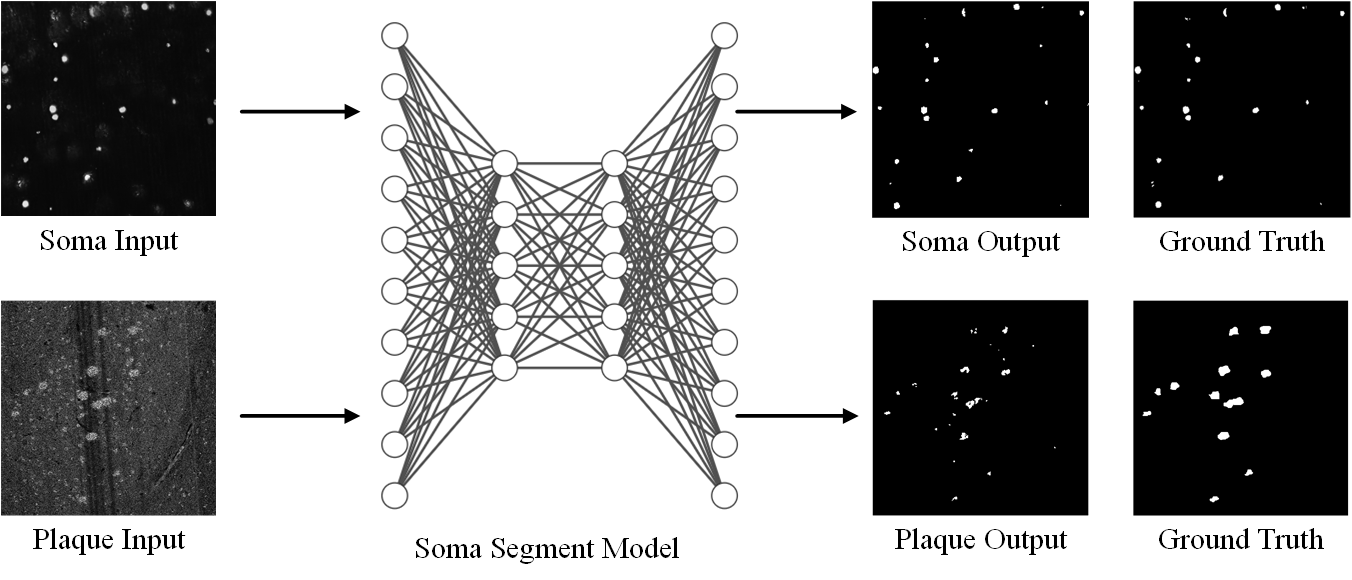}}
\caption{Similarity between segmentation tasks}
\label{fig:Similarity}
\end{figure}

As shown in Fig.~\ref{fig:Similarity}, neuron somas and cerebral cortex plaques in mouse brains are distinct biological structures,
but the segmentation task for them have the same characteristics:
both are grayscale images,
and target structures are concentrated areas with high grayscale values.
Due to the imperfection of sample preparation protocol,
a large number of false signals may be introduced in the imaging process,
the corresponding segmentation and identification labels will be difficult to obtain.
However, the labeling and imaging techniques for the neuron somata are relatively mature,
so it is not difficult to train a segmentation model for soma segmentation,
and due to the similarity of the image styles,
this model also has certain segmentation ability for cortical plaques.
Thus the segmentation results usually have a clean black style except for foreground areas,
we can make the denoised images close to this segmentation style,
so as to further improve the signal-to-noise ratio.
%For complex noise distribution, many noise signals will remain in the denoised image.
However, the result may still keep some noise as the downstreaming task is not exact the same as the target task.

The specific approach is that pre-train an available segmentation model on a similar task, such as the soma segmentation.
Since the segmentation branch provides an expected image style rather than pixel classification,
we prefer to use the MSE loss to pre-train and then cascade them after the denoising model.
The denoising model follows the same structure described in Sec.~\ref{subsec:se2e},
but the newly added branch calculates MSE loss with the result of segmentation directly.
In order to keep the expected image style unchanged,
the parameters of the segmentation model are fixed  in the whole training process.
The combined loss of unsupervised end-to-end training is shown in Equ.~\eqref{equ:unsupervisedLoss}
\begin{equation}
  l = w_1 l_1(F_{dn}(X)) + w_2 l_2(F_{seg}(F_{dn}(X)))
  \label{equ:unsupervisedLoss}
\end{equation}.

% TODO: move network structure and loss here.

\section{Experiment}
\label{sec:exp}
% TODO: use all the LM.
In this experiment, we provide the experiment result for the end-to-end networks on electron microscope
as well as optical microscope images
and compare the denoising performance against other baseline methods.
We first test the supervised end-to-end model in Sec.~\ref{subsec:data},
which is tested on the paired hippocampus mitochondrial segmentation images.
Then we test the unsupervised end-to-end model in Sec.~\ref{subsec:dmodel} on the optical microscopic slice images of rat brain.
We not only measure the ability of models to remove noisy signal,
but also investigate the benefit of denoised images for the downstreaming tasks.

\subsection{Data Description}
\label{subsec:data}
\begin{figure}[htbp]
\centerline{\includegraphics[width=0.49\textwidth]{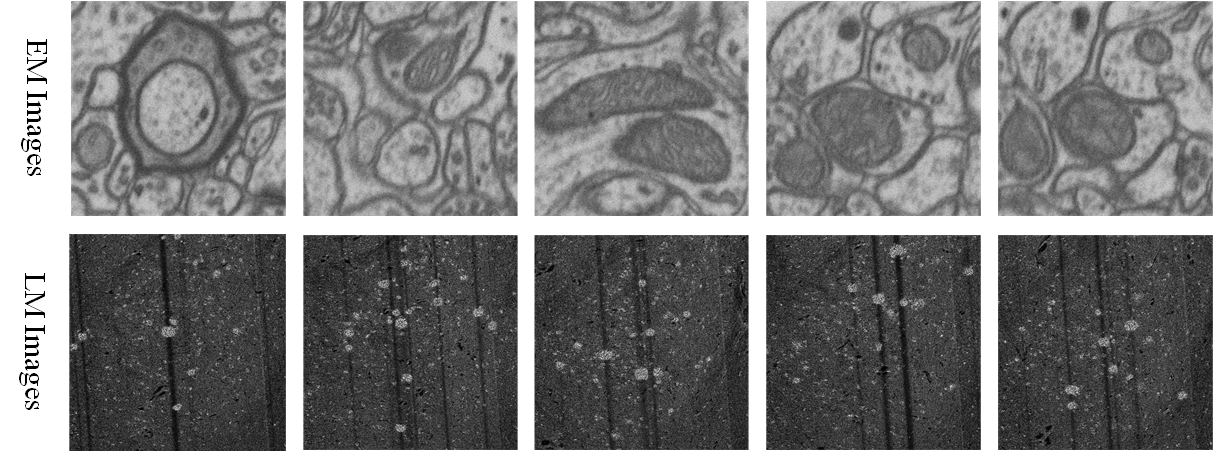}}
\caption{EM and LM images are used in experiments}
\label{fig:data}
\end{figure}

There are two images datasets(Fig.~\ref{fig:data}) used in the training and the prediction of the proposed end-to-end network.
Due to the fact that an self-supervised model is used as denoiser,
clean medical images as ground truth are not needed.
However, it is hard to ensure that downstream tasks could always be trained without paired training data, too.
Thus, these two situations should be discussed separately.

In this paper, supervised segmentation model is trained with a set of public EM mitochondrial segmentation dataset~\cite{EMdata},
which represents a $5 \times 5 \times 5 \mu m$ section taken from the CA1 hippocampus region of the brain.
The pixel size of original images is $1065 \times 2048$,
and we cut it into $256 \times 256$ tiles.
An enhancement procedure of rotation and flip is used to preprocess these images.
After the enhancement, $19488$ tiles are collected as the training set,
while the remaining $512$ tiles are used as the test set.
A generated Gaussian noise with different variance was added to the original images to make a noisy and clean image pair.

For the whole unsupervised end-to-end network,
we adopt the optical microscope slice images of the mouse brain containing $A\beta$ plaques.
$A\beta$ plaques are thought to be closely related to Alzheimer's disease. 
With the help of optical imaging systems, neuroscientists can image $A\beta$ plaques in mouse brain, so as to study their morphology and distribution.
However, the noise generated in the imaging process could seriously hinder the research.
Here we tested a set of mouse brain $A\beta$ plaques images,
and variation in the sample preparation process and defects of the imaging system
introduce the actual noise in this dataset,
which hinder the subsequent plaque recognition.
The imaging resolution is $0.32\mu m$ per pixel along both the x and y-axis and $ 1 \mu m$ per pixel along the z-axis.
The size of each images is $256 \times 256$ and $479$ images for the training while $50$ images for the testing.
The $A\beta$ plaque recognition model was trained on another cleaner but from the same source dataset.
% TODO: add two row of images, one row for EM, one row for Abeta.

\subsection{Experiment Models}
\label{subsec:dmodel}
There are several deep learning models used in both supervised and unsupervised training.
Tab.~\ref{tbl:models} shows some parameters of these models, including structure, loss function and application.
The denoiser of the two training strategies is a  modified Noise2Void model.
Original N2V model cuts input images into a series of 64×64 sub blocks,
% TODO: ?
% TODO: move to the method section.
%which could be hardly feed into connected segmentation model.
Differently, we input the full size image directly,
and divide it into dozens of small areas,
then mask the center pixel randomly for each small area.
The advantage of this is that denoised images are still full size
so can be input into subsequent models immediately.

The downstream segmentation model is an improved U-Net,
whereas batch normalization layers are inserted before almost every activation function.
In addition, we used leakey rectified linear unit (LReLu) as activation function instead of rectified linear unit (ReLU).
In order to form the final result, the original input is added to the network output, which allows the network only learns to extract the noise from the input, rather than the content.

The cross entropy loss function is used in the supervised segmentation model.
However, for unsupervised training,
the segmentation branch only provides a kind of image style for the denoiser to learn rather than doing pixels classification.
% TODO: move the loss funciton description to the method seciton.
Thus we choose the mean square error (MSE) loss function for the  unsupervised segmentation model.
Typical Faster-RCNN model was used in terms of plaque recognition task.

\begin{table}[htb]
\caption{Settings of Models}
\begin{tabular}{|c|c|c|c|}
\hline
Name                    & Improvement                                                                          & Loss Function                                                 & Purpose                                                          \\ \hline
                        & \begin{tabular}[c]{@{}c@{}}add BN layer,\\ Leakey ReLU,\\ residual unit\end{tabular} & n2v loss                                                      & denoiser                                                         \\ \cline{2-4} 
                        & \begin{tabular}[c]{@{}c@{}}add BN layer,\\ Leakey ReLU\end{tabular}                  & cross entropy                                                 & \begin{tabular}[c]{@{}c@{}}mitochondrion \\ segment\end{tabular} \\ \cline{2-4} 
\multirow{-3}{*}{U-Net} & \begin{tabular}[c]{@{}c@{}}add BN layer,\\ Leakey ReLU\end{tabular}                  & \begin{tabular}[c]{@{}c@{}}mean \\ squared error\end{tabular} & plaque segment                                                   \\ \hline
Faster R-CNN            & standard structure                                                                   & {\color[HTML]{4D4D4D} multi-task loss}                        & plaque detect                                                    \\ \hline
\end{tabular}
\label{tbl:models}
\end{table}

% move to here for better format
\begin{figure*}[htbp]
\center{\includegraphics[width=0.95\textwidth]{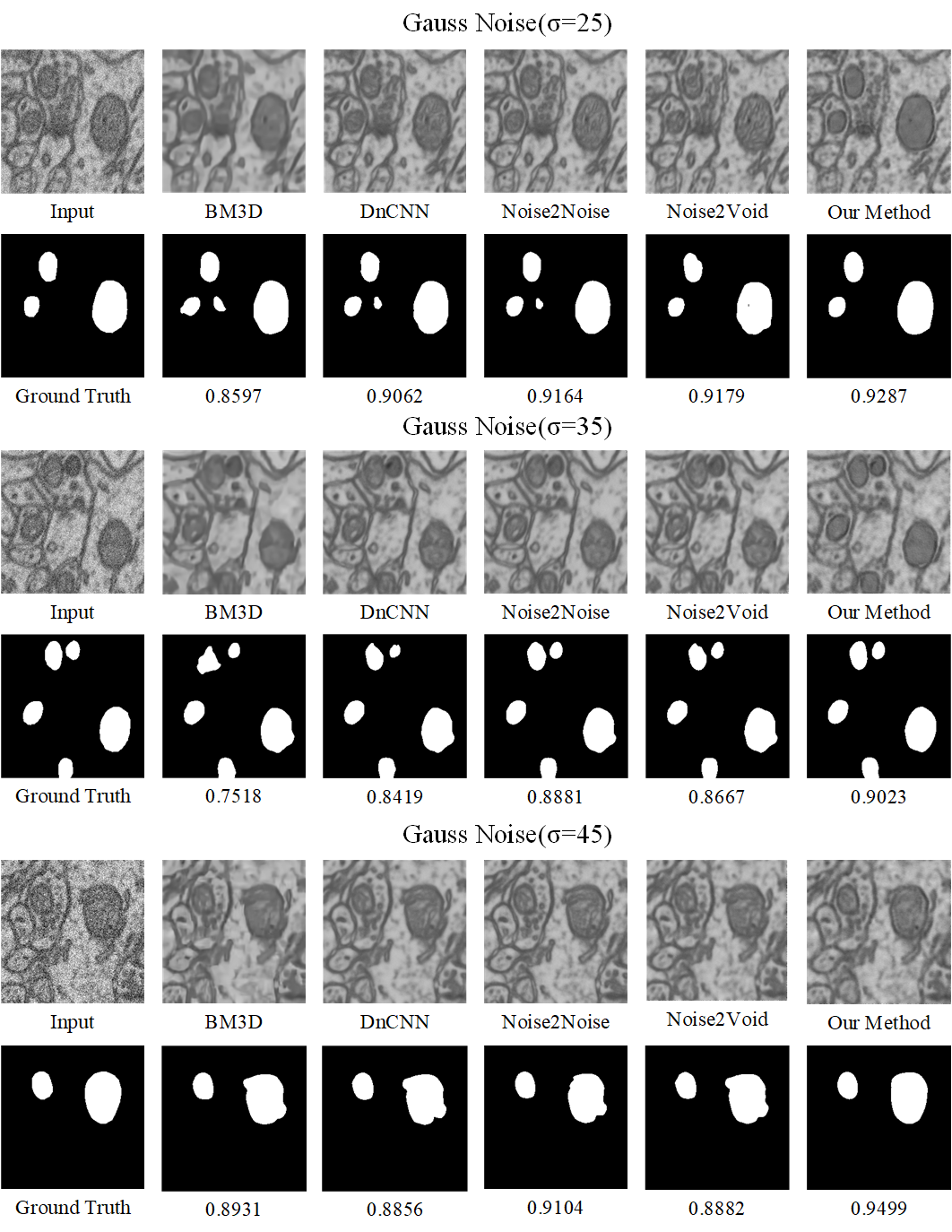}}
\caption{Comparison of denoising effect with different methods on EM images.}
\label{fig:exp:supervised}
\end{figure*}

\subsection{Experiment Setting}
\label{subsec:experiment}
This section is about some training details. 
We used Adam optimizer to train deep learning models and learning rates were selected from 0.0004 to 0.001. 
The weights of denoising loss and segmentation loss when training the supervised end-to-end model were 1 and 1.5, and they were 9 and 1 when training the unsupervised end-to-end model.

\begin{figure}[htb]
\centerline{\includegraphics[width=9cm]{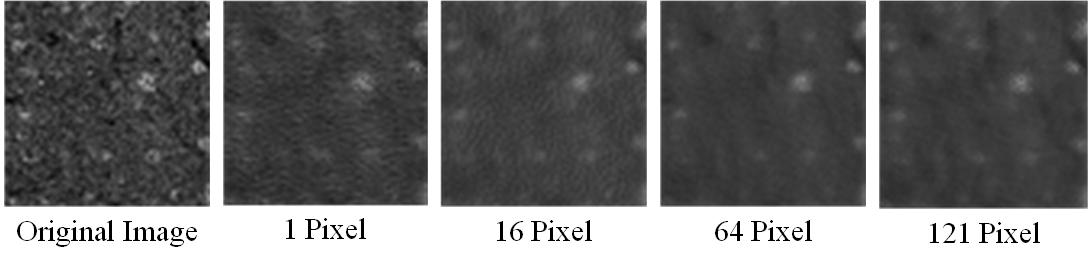}}
\caption{N2V performance with different masked pixels}
\label{fig:masked pixels}
\end{figure}

We point out in particular that Sec. ~\ref{subsec:dmodel} proposes a strategy of training N2V model directly from normal images and was used in the end-to-end model.
Fig. ~\ref{fig:masked pixels} shows how the effect changes as the number of masking pixels increases.
When too few pixels are masked, N2V model is hard to obtain a strong ability of denoising. 
But when the number of masked pixels exceeds a certain degree, the model would be stable in a finest place.

\begin{figure}[htb]
\centerline{\includegraphics[width=9cm]{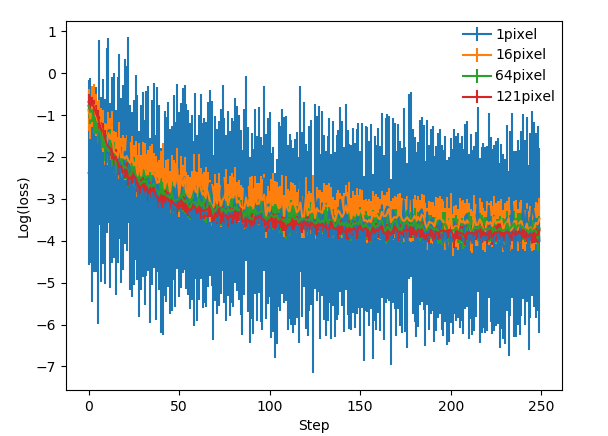}}
\caption{The training loss with a different number of masked pixels}
\label{fig:n2v lossine}
\end{figure}

Fig.~\ref{fig:n2v lossine} indicates the relationship between the training loss trend and the number of masking pixels and the ordinate is the logarithm of training loss.
We train each model ten times with random initialized parameters, and then calculated the total variance, reflected in the width of each curve. Therefore, the width of each curve at different steps is the same, which represents the average variance of all steps.
It is shown that when only one pixel is masked, the training loss decreases unsteadily.
As the number of masking pixels increases, the decreasing rate of the training loss will be slightly slower, but the convergence is significantly more stable.
Based on the experimental results above, we mask 169 pixels when training the end-to-end models.

\subsection{Model Supervised Soma Segmentation}
\label{subsec:somaseg}
We first test our supervised method over the EM mitochondrial dataset.
The comparison of denoising effects between different models is shown in Fig.~\ref{fig:exp:supervised}.
We set three different noise variances and tested the effect of mitochondrial segmentation on each denoised image.
The first column is the noisy image input from the EM dataset;
the second to fourth columns are the outputs of four different baseline denoising methods;
the last column is the image denoised by our end-to-end method. Under input noisy images are their corresponding mitochondrial labels.
Below the denoised images output by different methods is the performance
and Intersection-over-Union (IOU) score of mitochondrial segmentation with a same pre-trained segmentation model.

It is obvious that for the three different kinds of noise intensity,
the edge of mitochondria segmented by the proposed end-to-end method is the smoothest and the shape is the closest to ground truth.
Besides, we can find the denoiser of end-to-end method not only removes the Gaussian noise signal,
but also makes the mitochondria which neuroscientists focus on standing out from the image background,
separated from unimportant signals.
This is because the targets with prominent edge are more advantageous to segmentation model in the process of joint training,
and the denoiser has learned this demand.
%E2E column have not replaced

\begin{figure}[htbp]
\centering
\subfigbottomskip=2pt
\subfigure{
\includegraphics[width=0.47\linewidth]{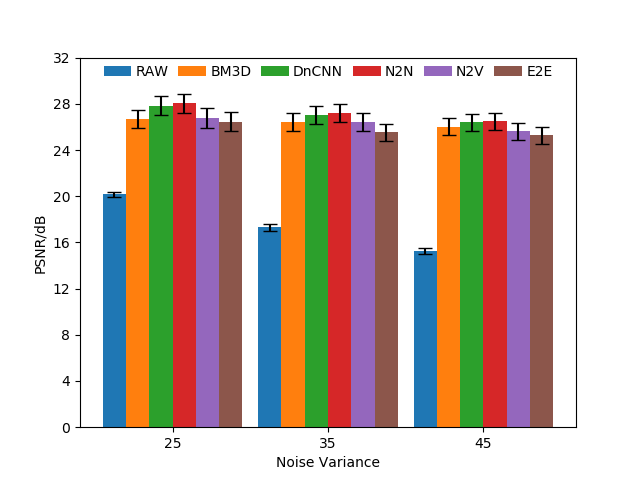}
}
%\quad
\hspace{-5mm}
\subfigure{
\includegraphics[width=0.47\linewidth]{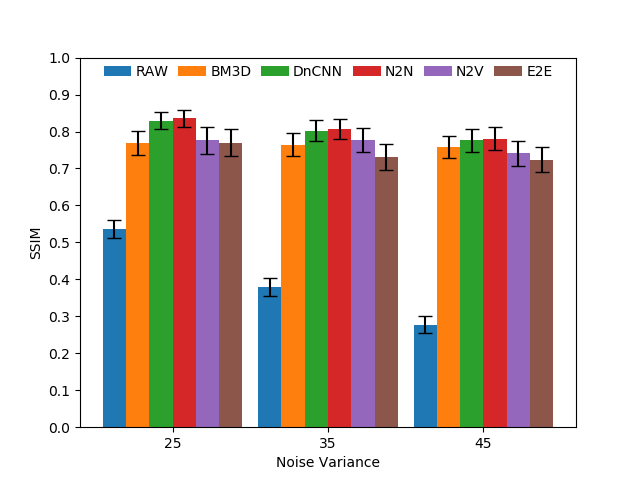}
}
%\quad
\subfigure{
\includegraphics[width=0.47\linewidth]{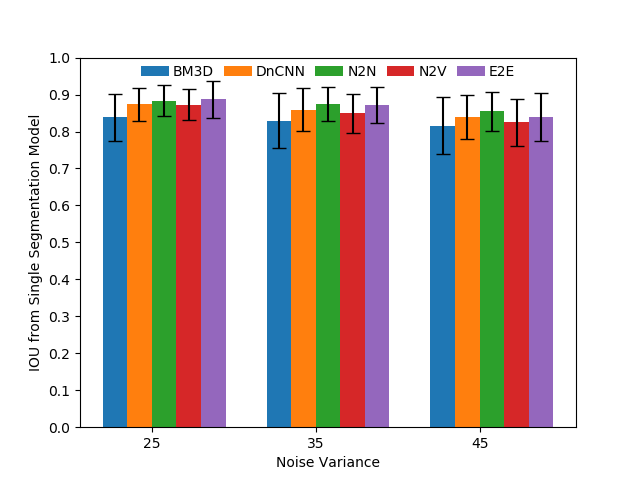}
}
%\quad
\hspace{-5mm}
\subfigure{
\includegraphics[width=0.47\linewidth]{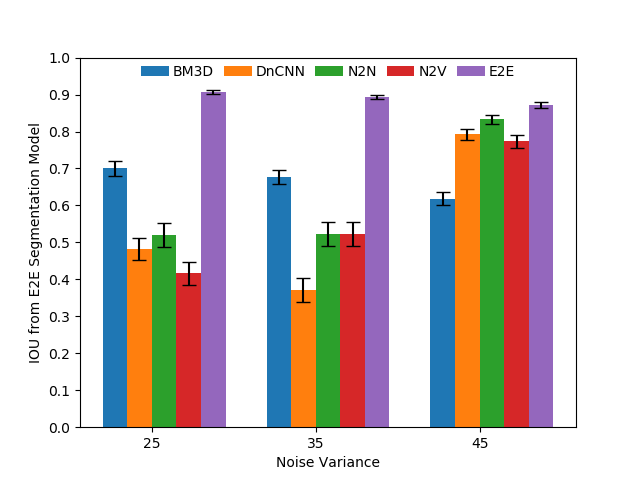}
}
\caption{Scores of different denoising models}
\label{fig:Barcharts}
\vspace{0.05in}
\end{figure}

Owing to that the end-to-end denoising model is affected by downstream tasks, its  output is no longer only consistant with the original image, which makes the PSNR and SSIM score not significantly higher than other methods as shown in Fig.~\ref{fig:Barcharts}. 
However, we do not believe that this subtle change will have a huge impact on specific medical image analysis, especially if it markedly promotes downstream tasks. 
As can be seen from the last two histogram of Fig.~\ref{fig:Barcharts}, 
the images denoised by end-to-end method show significant advantages on the joint trained segmentation model, and not lag behind on the separately trained segmentation model.  

\begin{figure}[htbp]
\centerline{\includegraphics[width=9cm]{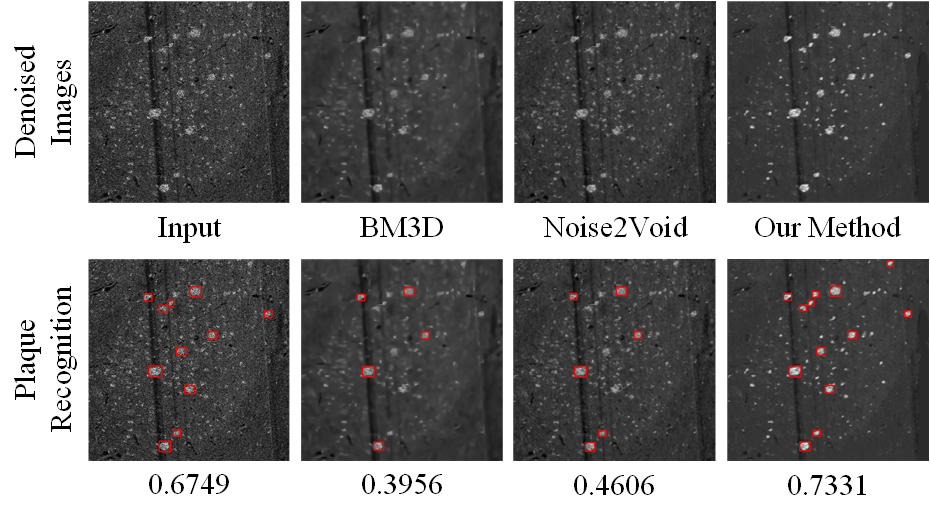}}
\caption{Comparison of denoising effect with different methods on LM images.}
\label{fig:exp:unsupervised}
%\end{figure*}
\end{figure}

\subsection{Label Supervised Amyloid Plaques Segmentation}
\label{subsec:plagseg}
We then present the unsupervised denoising and plaque recognition result on $A\beta$ LM images.
$A\beta$ plaque is a kind of important markers of Alzheimer's disease, 
and it will present various shapes during the different developing periods of mouse brains.
The noise generated in the imaging process will seriously hinder the brain scientists from analyzing the plaques, such as plaque recognition.
As shown in Fig.~\ref{fig:exp:unsupervised}, the images at the first row are the output images of three unsupervised denoising methods without ground truth,
while the second raw are the result and IOU scores of plaque recognition with a Faster-RCNN model using these images.

It can be seen that there are many dense noise signals all over the original $A\beta$ images.
% TODO: add more baseline, N2N, dnCNN...
However, BM3D and Noise2Void methods can only remove part of the noise,
and even change the original signal to a certain extent
because noise signals and plaque signals are closely entangled,
which reduces the accuracy of plaque recognition.
The output image background of end-to-end method is the cleanest thus Faster-RCNN model identified more plaques.  

\begin{table}[h]
\centering
\caption{IOU of plaque recognition with denoised images} % this is on test dataset
\scalebox{1.3}{
\begin{tabular}{|c|c|c|c|c|}
\hline
Image   & RAW    & BM3D   & N2V    & E2E    \\ \hline
1       & 0.6362 & 0.4454 & 0.3930 & 0.7218 \\
2       & 0.6749 & 0.3956 & 0.4606 & 0.7331 \\
3       & 0.6365 & 0.5644 & 0.5549 & 0.6419 \\
4       & 0.7440 & 0.7359 & 0.6958 & 0.8072 \\
5       & 0.6897 & 0.5043 & 0.5151 & 0.7172 \\
Average & 0.7169 & 0.5598 & 0.5173 & 0.7305 \\ \hline
\end{tabular}}
\label{tbl:IOU}
\end{table}

\begin{table}[h]
\centering
\caption{F1-Score of plaque recognition with denoised images}
\scalebox{1.3}{
\begin{tabular}{|c|c|c|c|c|}
\hline
Image   & RAW    & BM3D   & N2V    & E2E    \\ \hline
1       & 0.6897 & 0.4000 & 0.3478 & 0.8125 \\
2       & 0.7500 & 0.2222 & 0.4211 & 0.8462 \\
3       & 0.5556 & 0.7143 & 0.7143 & 0.7778 \\
4       & 0.8966 & 0.8571 & 0.8148 & 0.9333 \\
5       & 0.7556 & 0.5946 & 0.5946 & 0.8889 \\
Average & 0.8144 & 0.5519 & 0.5387 & 0.8211 \\ \hline
\end{tabular}}
\label{tbl:F1}
\end{table}

To be more precise, Tbl.~\ref{tbl:IOU} and Tbl.~\ref{tbl:F1} illustrate the explicit IOU and F1-Score of several plaque images,
whose labels are drawn by professionals.
The last row of the two tables represents the average score across the whole test dataset.
It could be found that although the quality of the plaque recognition results of original images are good
due to the strong robustness of Faster-RCNN model,
images processed by our method can still further improve the recognition performance, achieving higher IOU and F1-Score.

\section{Conlcusion}
\label{sec:conclusion}
In this paper, we put a step further on the biomedical image denoising
to tackle the question
of lacking denoising labels in the optical mesoscopic neuroimaging.
To answer the question, we look at the image denoising itself
and beyond to the following semantic tasks, such as segmentation and recognition.
The denoising model first starts with the Noise2void model as the backbond
for the optical mesoscopic image denoising.
Then the output of the Noise2void model feeds to the downstreaming semantic tasks
and the manually generated labels for the semantic tasks, 
forming an end-to-end network to
drive both the denoising model and the semantic task model
to give a better performance.

In general, due to the defects introduced during the sample preparation and the optical imaging,
the noise almost always present in the microscopic imaging results,
which hinders the neuroscientist from performaing the morphological analysis.
While the paired training data that is necessary in the supervised denoising models
is difficult to obtain from the microscopic imaging system alone.
Thus the unsupervised denoising models would have a wider application in neuroimaging.
Besides, the structures that are not the interest of the neuroscientists
could disturb the study of the concerned brain regions,
even if they are biological structures with their own functions.
Therefore, today's denoising methods are not only focusing on removing the noise in the images, 
but also expected to weaken the unimportant regions or enhance the targets of interest.

Based on the two observation, our end-to-end network uses
a self-supervised denoiser, and divide the model into two cases
according to the difficulty of downstream tasks to obtain paired training data.
We then apply our model in the experiments with EM and LM images respectively.
The results show that the end-to-end network can enhance the relevant signals
and weaken the uninterested signals according to the joint downstream tasks 
while steadily remove the noise.
Compared with other denoising models trained separately, 
the images processed by our method can obtain the highest score in mitochondrion segmentation and $A\beta$ plaque recognition.
In the follow-up research, 
we will focus on finding the common features of different end-to-end networks, 
so as to realize the fast replacement of denoisers.

\bibliographystyle{IEEEtran}
\bibliography{./qhe}

\end{document}